Perspective

# Ultrafast Sintering

**Jian Luo** (✉)

Aiiso Yufeng Li Family Department of Chemical and Nano Engineering, Program in Materials Science and Engineering, University of California San Diego, La Jolla, CA 92093, USA

## GRAPHICAL ABSTRACT

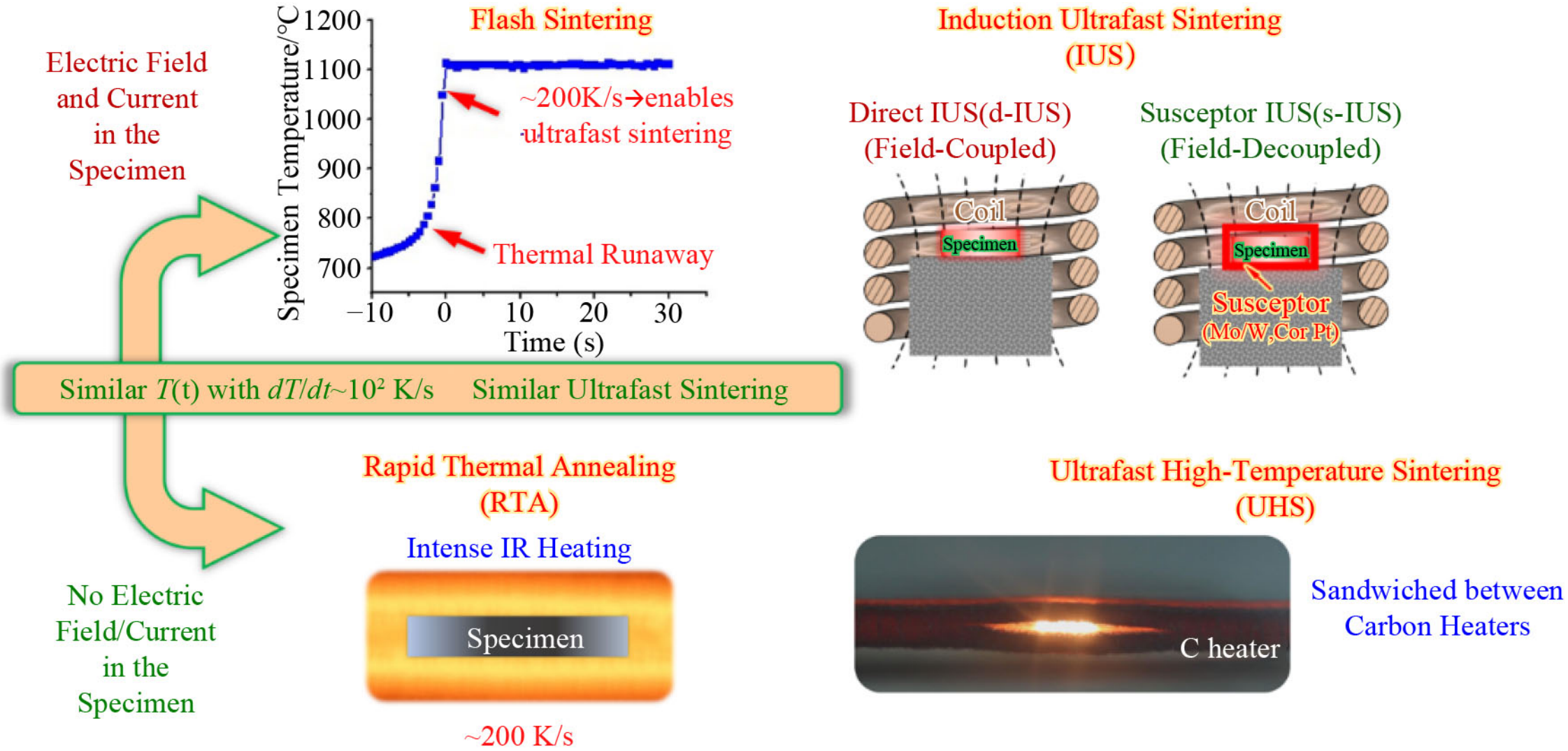

## HIGHLIGHTS

- Various ultrafast sintering schemes with and without electric fields are developed.
- Ultrahigh heating rates and high sintering temperatures enable rapid densification.
- Reactive ultrafast sintering of compositionally complex ceramics is achieved.
- Ultrafast sintering methods enable high-throughput materials discovery.
- Ultrafast sintering kinetics and mechanisms require further investigation.

**ABSTRACT** This Perspective critically assesses recent advances in ultrafast sintering and highlights open scientific questions and emerging technological opportunities. Mechanistic studies of flash sintering indicate that the flash event initiates as a coupled thermal and electrical runaway, while rapid densification is enabled by ultrahigh heating rates and elevated sintering temperatures. Building on this understanding, ultrafast sintering has been realized without passing electric currents through the specimens via multiple approaches, including rapid thermal annealing (using intense infrared heating), ultrafast high-temperature sintering (in which specimens are sandwiched between graphite felt heaters), blacklight sintering (employing blue laser or intense ultraviolet irradiation), atmospheric-pressure plasma sintering, and induction ultrafast sintering (utilizing skin currents in direct induction heating or no current in the specimens in susceptor-heating mode). Reactive ultrafast synthesis and sintering have also been demonstrated. Although several hypotheses have been proposed, the mechanisms governing ultrafast sintering and its kinetics warrant further investigation. In particular, reactive ultrafast synthesis and sintering of compositionally complex ceramics are scientifically intriguing to understand while also presenting technological opportunities. The expanding range of ultrafast sintering methods provides a versatile platform for high-throughput materials discovery, especially in the rapidly growing field of high-entropy and compositionally complex ceramics, which feature vast compositional spaces to explore.

**KEYWORDS** Ultrafast sintering; Flash sintering; Ultrafast high-temperature sintering; Induction ultrafast sintering; Reactive ultrafast sintering; Sintering

✉ Corresponding author. E-mail: jluo@alum.mit.edu

# 1 Introduction

Ceramic manufacturing via sintering, a milestone of human civilization dating back over 26,000 years [1], often requires firing at high temperatures (*e.g.*, 800–2000 °C) for hours. This process consumes substantial energy, generates significant $CO_2$ emissions, and is costly. Early studies of "fast firing" have been noted [2,3]. Spark plasma sintering (SPS), more appropriately called field-assisted sintering technology (FAST), has attracted considerable interest because it can achieve high densification at lower temperatures and in shorter times compared with conventional sintering and hot pressing (Fig. 1) [4].

Recently, significant efforts have focused on developing innovative sintering technologies to reduce energy consumption [5]. Two groundbreaking examples are "cold sintering" [6−8] and "flash sintering" [9]. In a further example, "water-assisted flash sintering" used water vapor to trigger the flash of a ZnO green specimen at room temperature, achieving >98% relative density without a furnace [10]. A subsequent study reported the development of a similar scheme of "flash cold sintering" [11]. Other innovative sintering technologies have been demonstrated, including electro-sintering (driven by applied electric currents) [12−14] and various ultrafast methods that do not rely on electric fields or currents, such as rapid thermal annealing (RTA) [15], ultrafast high-temperature sintering (UHS) [16], and induction ultrafast sintering (IUS) [17]. Figure 1 illustrates selected sintering technologies as a function of furnace temperature and sintering time. This perspective article critically assesses advances in ultrafast sintering.

# 2 Flash Sintering

In 2010, Professor Raj and co-workers invented flash sintering [9], using an applied electric field to trigger a "flash" and rapidly densify 3 mol.% $Y_2O_3$-stabilized $ZrO_2$ (3YSZ) in ~5 seconds at a nominal furnace temperature of 850 °C. This work [9] immediately attracted significant scientific and technological interest, and researchers worldwide soon extended flash sintering to a wide range of ceramic materials [18−21]. Flash sintering differs from SPS/FAST in that it employs higher electric fields, shorter sintering times, and lower furnace temperatures. Notably, specimen temperatures are typically much higher due to Joule heating.

Three key scientific questions are:

How does the flash initiate?

What enables ultrafast densification?

How do electric fields influence sintering and microstructural evolution?

Different mechanisms have been proposed to explain flash sintering. On one hand, Raj et al. suggested that flash sintering involves an avalanche generation of point (Frenkel) defects [21,22], which can produce electroluminescence [22−24], and that the flash is initiated at a critical power density [25]. However, Schie et al. showed that the electric field required to generate a Frenkel pair is as high as ~$10^{10}$ V/m for $HfO_2$ [26]. Raj et al. further proposed that proliferation of phonons at the edge of the Brillouin zone can induce Frenkel pairs even without an applied electric field [27].

On the other hand, several studies have suggested that a flash often begins as a thermal runaway [28−30], with ultrahigh heating rates promoting ultrafast densification [15,31], as discussed below.

## 2.1 How does a flash start?

In 2015, Luo and co-workers (my research group) reported that a flash can initiate as a coupled thermal–electrical runaway [28]. This arises from the Arrhenius dependence of the specimen conductivity, leading to thermal runaway when the increase in Joule heating exceeds the corresponding increase in radiative heat dissipation with rising temperature [28]. A quantitative model Fig. 2(c) [28] was proposed using measured temperature-dependent conductivities to predict the thermal runaway temperatures, which agreed well with the observed onset temperatures of flash sintering in ~20 different cases [15,18,28,32,33]. Similar thermal runaway models have been independently proposed by Todd et al. [29] at the same time and by Dong and Chen [30,34] slightly later, and validated for additional material systems.

However, the success of thermal runaway models does not rule out the possibility that a sudden increase in specimen conductivity caused by other physical processes can also trigger a flash as a "forced" runaway, in addition to the natural thermal runaway arising from the Arrhenius-like increase of conductivity with temperature. For example, flash sintering can be activated by the formation of an ion-conducting eutectic liquid or premelting-like interfacial liquid in $Bi_2O_3$-doped ZnO, which induce an abrupt increase in specimen conductivity [35].

## 2.2 What enables ultrafast sintering?

To identify the key factors enabling ultrafast sintering, Luo and co-workers (my research group) reported a critical mechanistic study in 2017, demonstrating that similar ultrafast densification of ZnO can be achieved via (1) flash sintering (with an applied electric field) and

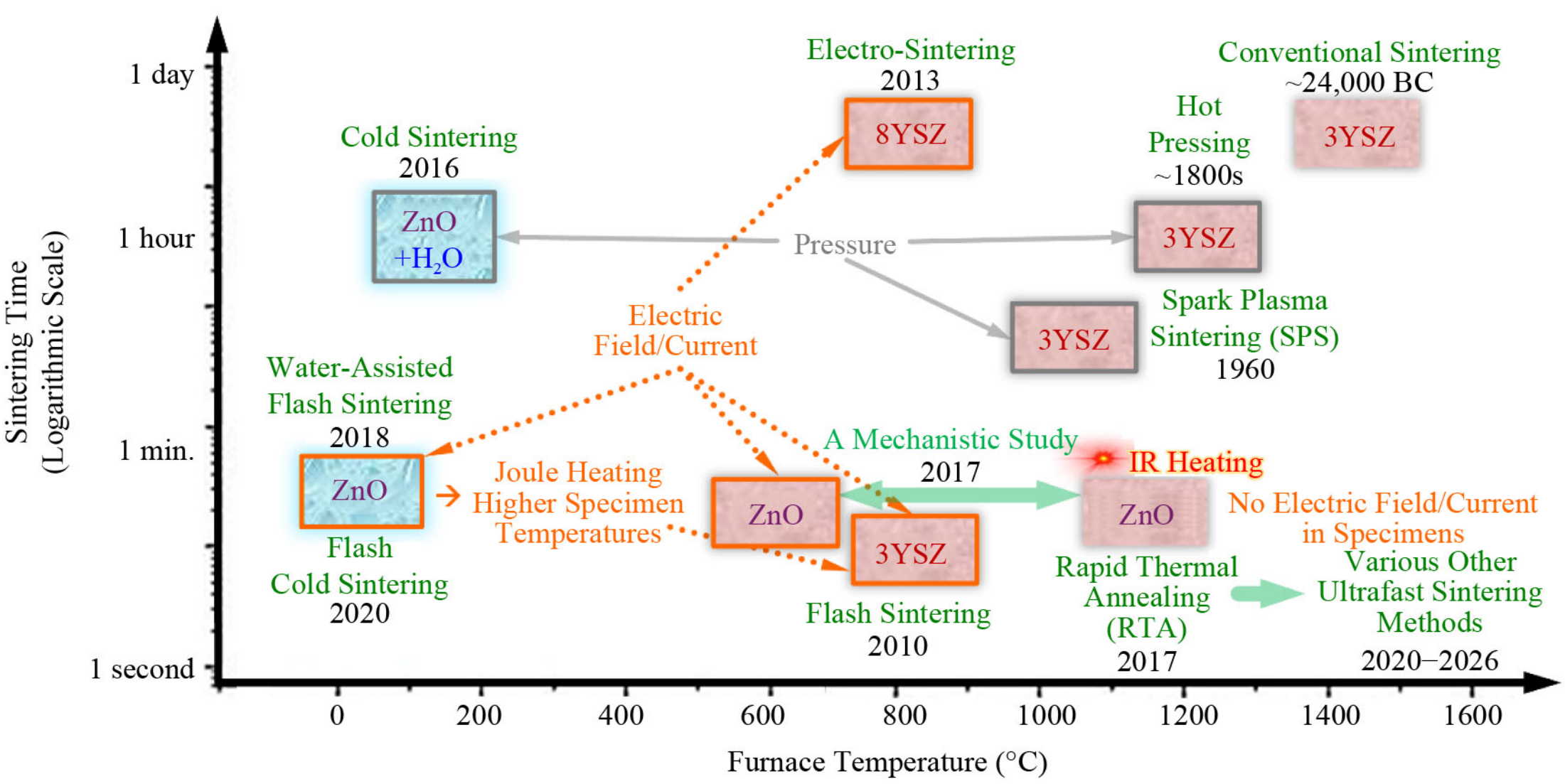

**Fig. 1** Comparison of selected sintering technologies plotted as a function of sintering time and furnace temperature, using 3 mol% and 8 mol% yttria-stabilized zirconia (3YSZ and 8YSZ) and ZnO as representative examples. This figure is based on a comparison of flash sintering [9,21] with conventional sintering, hot pressing, and spark plasma sintering (SPS), as illustrated in an online news image by Marco Cologna and Rishi Raj, where the author has further incorporated cold sintering [6−8], electro-sintering [12−14], water-assisted flash sintering [10], flash cold sintering [11], and rapid thermal annealing (RTA) [15] for comparison.

(2) RTA (or rapid thermal annealing using infrared heating without an electric field; Fig. 2(d)) when subjected to similar temperature profiles with heating rates of ~200 °C/s (Fig. 2(b)) [15]. This work further showed that flash sintering and RTA produced comparable densification curves (Fig. 2(e)) and grain growth kinetics, and densification was reduced when the heating rate was lowered [15]. Thus, this study [15] indicates that high heating rates ($dT/dt$) on the order of $10^2$ K/s, not electric fields, are the primary driver of ultrafast sintering [15]. In the same year, Ji et al. (Todd and co-workers) independently reported an ultrafast firing experiment using an exothermic Ni–Al reaction and reached a similar conclusion for 3YSZ [31].

While general characteristics of ultrafast sintering have been identified, differences in sintering kinetics, as well as in microstructural evolution and defect formation, across various material systems, which can often be material-specific, require case-by-case investigation.

## 2.3 Electric field effects

Although the prior study demonstrated that ultrafast sintering can be achieved without an applied electric field (*e.g.*, via RTA using infrared heating; Fig. 2(d)−(e)) [15], this does not preclude the possibility that electric fields can influence sintering kinetics and microstructural evolution.

Notably, Chen and co-workers demonstrated that ionomigration of pores can induce "electro-sintering" of 8YSZ at temperatures hundreds of degrees below conventional sintering conditions (Fig. 1) [12−14]. This case provides an unambiguous example of an electric current effect enhancing sintering kinetics (albeit not in the ultrafast regime). However, the generality of this mechanism across other materials systems remains unclear.

Several studies have further suggested that applied electric fields or currents can modify (often accelerate) interfacial reaction kinetics, including $TiO_2$-$Al_2O_3$ [36], MgO-$In_2O_3$ [37], and $Al_2O_3$-spinel [38], and element-element interfaces [39−45]. Such field-induced enhancements in reaction kinetics could directly impact reactive ultrafast sintering processes, as discussed below.

In addition, a spectrum of intriguing electric-field effects has been reported, including both suppressed [46,47] and enhanced [28,35,46−50], and in some cases abnormal [48,49], grain growth. Asymmetrical grain growth has also been observed in flash-sintered $ZrO_2$ [51], 3YSZ [52], $MgAl_2O_4$ [53], $UO_2$ [54], and ZnO [28], among other systems.

In a further investigation of the underlying mechanisms through controlled long-annealing experiments, Luo and co-workers (my research group) demonstrated that an applied electric field can induce a grain boundary (GB) transition via electrochemical coupling, thereby altering grain growth [55]. We subsequently showed that applied electric fields can be used to controllably create and manipulate graded microstructures in ZnO [56] and $BaTiO_3$ [57]. In $BaTiO_3$, a cathode-side reduction-induced GB disordering transition at a higher applied voltage can further shift the region of enhanced grain growth from the anode to the cathode [57].

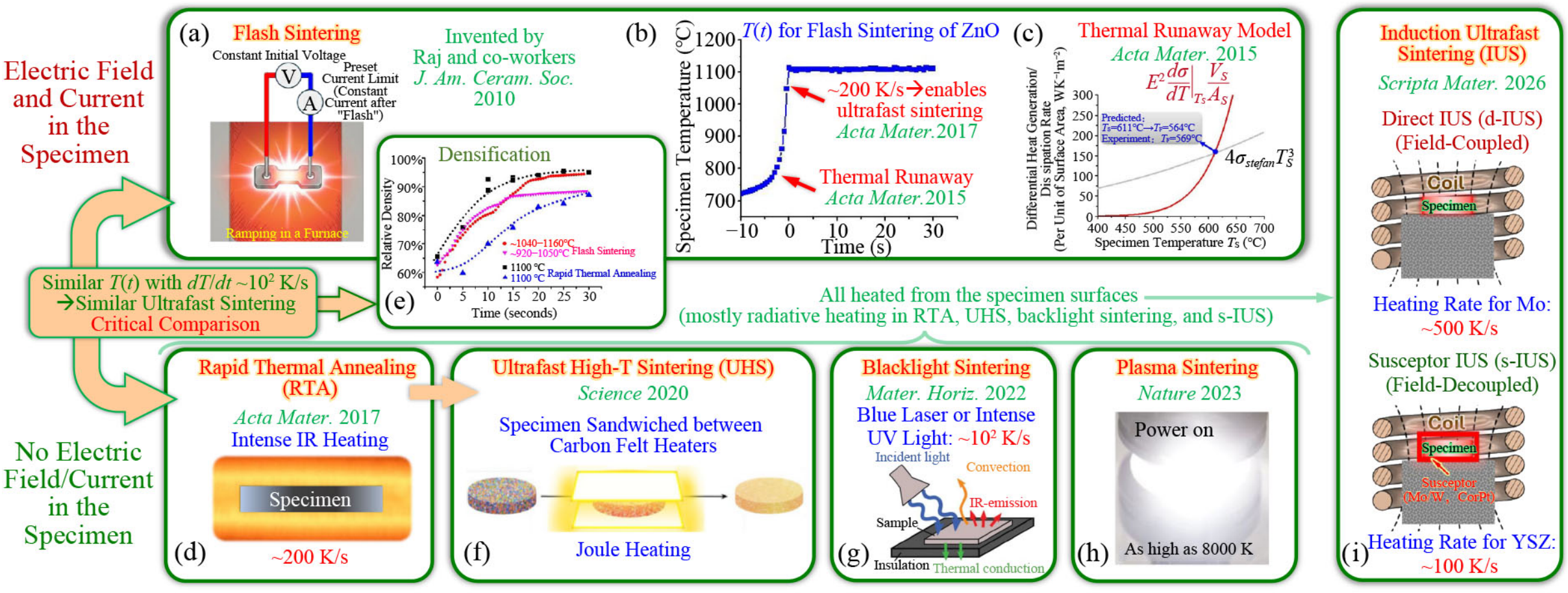

**Fig. 2** From a mechanistic study of flash sintering to a spectrum of ultrafast sintering without an electric field. (a) *Flash sintering*, invented by Raj and co-workers, uses an applied electric field to initiate sintering of yttria-stabilized zirconia (YSZ) within seconds [9]. (b) Temperature profile, $T(t)$, for a representative flash-sintering experiment on ZnO, showing an ultrahigh heating rate of ~200 K/s [15]. (c) In 2015, Luo and co-workers (my research group) proposed that a "flash" can originate from thermal runaway and developed a coupled thermal–electric runaway model that quantitatively predicts flash-onset temperatures [28], along with two other independent studies [29,30,34]. (d) The 2017 critical mechanistic study from my research group demonstrated that *rapid thermal annealing* (*RTA*) using intense infrared (IR) heating can achieve similarly ultrahigh heating rates to (e) produce comparable ultrafast densification without applying an electric field/current to the specimen [15]. This comparative study of ZnO ultrafst sintering with and without electric fields/currents [15], together with an independent ultrafast-firing study of YSZ [31], showed that ultrahigh heating rates of ~$10^2$ K/s can promote ultrafast densification without an applied electric field/current. Similar mechanisms enable (f) *ultrafast high-temperature sintering* (*UHS*), in which a sample is heated between two carbon felts serving as resistive heaters [16]; (g) *blacklight sintering*, which uses a blue laser or intense ultraviolet (UV) light for heating [63]; and (h) *atmospheric-pressure plasma sintering* [64], all featuring ultrahigh heating rates of ~$10^2$ K/s. Notably, in RTA, UHS, and blacklight sintering, the specimens are primarily heated at their surfaces through radiative heat transfer or by direct irradiation with electromagnetic waves. (i) A most recent study [17] from Luo and co-workers (my research group) reported a facile *induction ultrafast sintering* (*IUS*) approach that can operate in both field-coupled direct (d-IUS) and field-decoupled susceptor-heating (s-IUS) modes, mimicking flash sintering and UHS to some extent, respectively. Flash sintering schematic adapted from a base image generated by Gemini AI and further completed by the author. Other panels are adapted, with permission or under open-access licenses, from Zhang et al. [15] (© 2017, Elsevier), Wang et al. [16] (© 2020, AAAS), Porz et al. [63] (© 2022, CC BY 3.0), Xie et al. [64] (© 2023, Authors), and Shivakumar et al. [17] (© 2025, CC BY).

Prior studies have also suggested that electric fields can induce the formation of defects, such as point defects, dislocations, and stacking faults, during flash sintering, which may influence sintering kinetics, microstructural evolution, and mechanical and other properties [58–62].

It should be noted that electric-field–driven phenomena represent a broader topic, with numerous theories and observations. A comprehensive review of these phenomena is beyond the scope of this short perspective.

## 3 Ultrafast Sintering without Electric Fields in the Specimen

The 2017 mechanistic study [15] from Luo and co-workers (my research group) suggested the possibility of decoupling the electric response of a specific material from the processing protocol to develop more general ultrafast sintering technologies. The basic feasibility of this concept was already demonstrated by the RTA experiment reported in our 2017 study (Fig. 2(d)−(e)) [15]. Subsequently, ultrafast high-temperature sintering (UHS) was introduced as a general method to synthesize and densify ceramics within seconds by heating a thin specimen sandwiched between two carbon felts that serve as resistive heaters (Fig. 2(f)) [16].

Further extensions of ultrafast sintering without an applied electric field include blacklight sintering, which heats the sample using a blue laser or intense ultraviolet (UV) light (Fig. 2(g)) [63], and atmospheric-pressure plasma sintering [64], both achieving ultrahigh heating rates of approximately $10^2$ K/s. Here, it is noted that prior studies have reported plasma formation during flash sintering of boron carbide [65] and have utilized plasma electrodes for contactless flash sintering [66,67]; however, the effects of plasma on sintering (beyond heating) remain to be further investigated. Notably, in RTA, UHS, and blacklight sintering, the specimens are primarily heated at their surfaces through radiative heat transfer or by direct irradiation with electromagnetic waves. In addition, $CO_2$ and Nd:YAG lasers have been used to sinter optical [68,69], ferroelectric [70], dielectric [71], and structural [72−75] ceramics to high relative densities within seconds.

A most recent study [17] from Luo and co-workers (my research group) further introduced a facile induction ultrafast sintering (IUS; Fig 2(i)) approach that can operate in both field-coupled direct (d-IUS) and field-decoupled susceptor-heating (s-IUS) modes.

Here, a key advantage of ultrafast sintering technologies that involve no electric currents in the specimen is the decoupling of the specimen's electrical response, enabling the design of more general ultrafast sintering approaches without the need to consider material-specific electrical characteristics.

## 4 Reactive Ultrafast Sintering and Compositionally Complex Ceramics

Reactive ultrafast sintering, in which synthesis and densification occur simultaneously, is of particular interest and deserves further discussion. Reactive flash sintering has been applied to a wide range of materials, including (but not limited to) $MoSi_2$ [76], $BiFeO_3$ [77−80], (Ba, Sr)$TiO_3$ [81], $ZrTiO_4$ [82], $MgAl_2O_4$ [83−85], $Li_{5.95}Al_{0.35}La_3Zr_2O_{12}$ [86], $Li_{6.25}Al_{0.25}La_3Zr_2O_{12}$ [87], $Li_{0.5}La_{0.5}TiO_3$ [88], $LiNi_{1/3}Co_{1/3}Mn_{1/3}O_2$ [89], $Na_{0.5}K_{0.5}NbO_3$ [90,91], $Bi_{2/3}Cu_3Ti_4O_{12}$ [92], $MgAl_2O_4$-8YSZ [93], $Al_2O_3$-$Y_3Al_5O_{12}$ [94,95], $Ti_{0.5}Zr_{0.5}N$, and $Ti_{0.5}Al_{0.5}N$ [81]. Earlier studies of reactive flash sintering have been reviewed by Chaim et al. in 2021 [96] and Yoon et al. in 2023 [97]. Furthermore, reactive UHS has been applied to fabricate Ta-doped $Li_{6.5}La_3Zr_{1.5}Ta_{0.5}O_{12}$ [16], $Li_{1.5}Al_{0.5}Ge_{1.5}P_3O_{12}$ [98], and other oxide solid electrolytes [99].

Notably, ultrafast sintering has been employed to fabricate high-entropy ceramics (HECs) and compositionally complex ceramics (CCCs) (see two Perspectives [100,101] for further discussions of the emerging field of HECs and CCCs), primarily through reactive ultrafast sintering. Figure 3 illustrates two examples of ultrafast sintering of high-entropy borides, some of the most challenging ceramics to fabricate. In one case, a reactive SPS approach was used to simultaneously synthesize and sinter $(Hi_{0.2}Zr_{0.2}Hf_{0.2}Nb_{0.2}Ta_{0.2})B_2$ from a mixture of five binary borides in two minutes by flowing an electric current of 900 A through the specimen, resulting in a single-phase high-entropy boride with greater than 99% relative density (Fig. 3(a)) [102]. In another study, rapid liquid-phase–assisted ultrafast UHS was employed to fabricate a $(Ti_{0.2}Zr_{0.2}Ta_{0.2}Mo_{0.2}W_{0.2})B_2$-based composite in two minutes without applying electric current directly to the specimen (Fig. 3(b)) [103]. This approach used a composite powder pre-synthesized via a self-propagating exothermic reaction of elemental powders, so the ultrafast sintering in this case was largely non-reactive, although a liquid phase formed during the process and helped assist densification [103].

In addition, reactive flash sintering has been used to fabricate a wide range of HECs and CCCs, including $(Mg_{0.2}Ni_{0.2}Co_{0.2}Cu_{0.2}Zn_{0.2})O$ [104−108] and other [109] high-entropy rocksalt oxides, $Sr(Ti_{0.2}Y_{0.2}Zr_{0.2}Sn_{0.2}Hf_{0.2})O_{3-\delta}$ [110], $(Bi_{0.2}Na_{0.2}K_{0.2}Ba_{0.2}Ca_{0.2})TiO_3$ [111], and other [112] high-entropy perovskite oxides, (non-equimolar, medium-entropy) compositionally complex perovskite proton conductor $BaZr_{0.1}Ce_{0.7}Y_{0.1}Yb_{0.1}O_{3-\delta}$ [113], and several Li-containing high-entropy oxides [114−117], as well as high-entropy rare earth zirconates [118−121], ferrites [122−124], and nitrides [125].

In parallel, reactive UHS has been used to fabricate $(La_{0.2}Nd_{0.2}Sm_{0.2}Eu_{0.2}Gd_{0.2})_2Zr_2O_7$ [126], other medium-entropy [127] and high-entropy [128−131] oxides, high-entropy or compositionally complex carbides [132–135] (including thermally insulating porous carbides [135]), diborides and diboride-based composites [136–138], nitrides and carbonitrides [139,140], and refractory alloy–carbide composites [141].

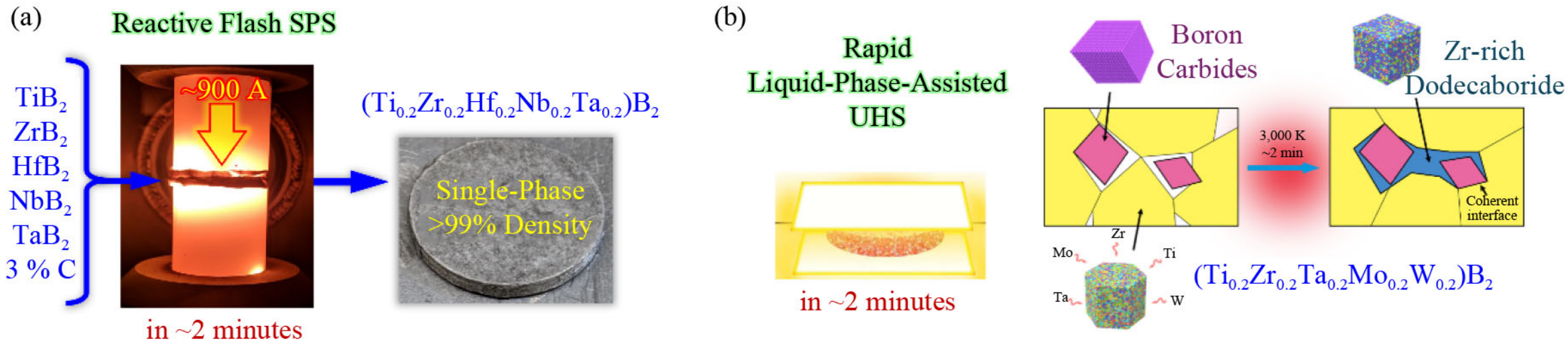

**Fig. 3** Ultrafast sintering of high-entropy borides, representing some of the most difficult-to-sinter ceramics. (a) Reactive ultrafast synthesis and sintering of $(Ti_{0.2}Zr_{0.2}Hf_{0.2}Nb_{0.2}Ta_{0.2})B_2$ from a mixture of five binary borides in ~2 minutes (with ~900 A electric current passing through the specimen) [102]. (b) Rapid liquid-phase–assisted ultrafast high-temperature sintering (UHS) of a $(Ti_{0.2}Zr_{0.2}Ta_{0.2}Mo_{0.2}W_{0.2})B_2$-based composite in ~2 minutes (without an electric current in the specimen), using a composite powder pre-synthesized via a self-propagating exothermic reaction of elemental powders [103]. Schematics and images were adapted, with permission or under an open-access license, from Gild et al. [102] (© 2017, Elsevier), Wang et al. [16] (© 2020, AAAS), and Xie et al. [103] (© 2022, CC-BY).

Moreover, ultrafast pressure-assisted sintering has been employed to fabricate high-entropy carbides [142,143], enabling accelerated materials discovery [143].

## 5 Mechanisms and Outlook

Ultrafast sintering has been realized through flash sintering (direct Joule heating), RTA (intense infrared heating), UHS (carbon felt heaters), blacklight sintering (blue laser or intense UV irradiation), plasma sintering, and IUS (direct induction or susceptor heating). Here, I hypothesize that common mechanisms underly these diverse approaches, promoted by ultrahigh heating rates and largely independent of the specific mode of energy delivery. Several possible mechanisms can be considered, as outlined in an earlier viewpoint article of flash sintering [18].

First, ultrahigh heating rates can suppress particle and pore coarsening, thereby promoting densification by maintaining high sintering driving forces. Particle and pore coarsening are often governed by surface diffusion, which has a relatively low activation energy and is more pronounced at intermediate temperatures (during heating ramping). By rapidly traversing this temperature regime, ultrafast heating can limit coarsening. This interpretation is supported by recent studies reporting smaller grain and pore sizes in ultrafast-sintered specimens compared with conventionally sintered counterparts [144–147]. Future work should quantitatively examine the kinetics of ultrafast sintering as a function of initial particle size and heating rate to more rigorously evaluate this proposed mechanism.

Second, it has been proposed that ultrahigh heating rates may lead to the formation of non-equilibrium GBs with enhanced diffusivities [15,18,31,147,148]. A model experiment showed that pre-sintering reduces the densification rate during UHS of alumina [139], providing indirect support for this hypothesis [147]. Additional evidence may come from reports of reduced apparent activation energies in ultrafast sintering [148]. Further studies are required to more directly probe GB structure and transport in order to critically assess this mechanism, although such investigations remain experimentally challenging.

Third, ultrafast sintering typically occurs at significantly higher temperatures but for much shorter durations. Under such conditions, premelting-like [149] disordered GBs may form, exhibiting markedly enhanced diffusivities beyond those expected from simple Arrhenius extrapolation. This proposed mechanism aligns with the broader concept of GB phase-like transitions [150–152], involving the formation of equilibrium or metastable GB phases (complexions). It is conceptually analogous to activated sintering [153–160], but occurring at higher temperatures through intrinsic GB premelting rather than segregation-induced GB disordering [153–160]. Hypothesized disordered GBs were observed by aberration-corrected scanning transmission electron microscopy in an unpublished collaborative study of YSZ. Future studies should directly characterize GBs in well-quenched or *in-situ* specimens and quantitatively analyze the associated kinetics to evaluate this hypothesis.

Furthermore, the mechanisms and kinetics underpinning reactive ultrafast sintering, particularly for CCCs, are likely more intricate than those in conventional ultrafast sintering. Reactive ultrafast processing involves the simultaneous occurrence of rapid synthesis, densification, and microstructural evolution, leading to coupled kinetic pathways. Here, I propose that a critical comparison between capillarity-driven conventional ultrafast sintering and entropy-driven versus enthalpy-driven reactive ultrafast sintering routes will provide valuable insights into both the general underlying mechanisms and the distinct kinetic pathways involved. Current studies are in progress to critically examine and compare these kinetic pathways for ultrafast sintering.

Various ultrafast sintering methods enable high-throughput materials discovery, particularly for HECs and CCCs, which have vast compositional spaces to explore.

Scaling up pressureless ultrafast sintering to produce large or complex-shaped components remains a significant technological challenge. Ultrafast sintering may first be scaled up for fabricating membranes or coatings with careful process control. Temperature gradients are likely to arise in thick specimens during large-scale UHS or other ultrafast sintering methods that rely on surface heating. Homogeneity must also be better managed in large-scale flash sintering, even though it involves volumetric heating. Pressure-assisted ultrafast sintering can help mitigate some challenges, but it also adds cost and processing complexity (and time).

Ultrafast sintering can modify phase transformation pathways and microstructural evolution, making it possible to stabilize non-equilibrium phases and microstructural features. Reactive ultrafast sintering, in particular, can promote the formation of non-equilibrium phases, interfaces, and defects, especially under large chemical driving forces, which can be further enhanced or alternatively facilitated by applied fields or other external stimuli. These possibilities create new technological opportunities for designing and fabricating non-equilibrium materials through ultrafast synthesis and sintering.

## 6 Conclusions

Ultrafast sintering has been realized through multiple approaches, including flash sintering (Joule heating with thermal runaway), rapid thermal annealing (intense infrared heating), ultrafast high-temperature sintering (graphite felt heaters), blacklight sintering (blue laser or intense ultraviolet irradiation), atmospheric-pressure plasma sintering, and induction ultrafast sintering. Reactive ultrafast sintering further combines simultaneous synthesis and densification in a single ultrafast process. Mechanistic studies indicate that rapid densification is primarily enabled by ultrahigh heating rates and elevated sintering temperatures. The underlying mechanisms require further investigation despite several proposed hypotheses. The growing array of ultrafast sintering methods offers powerful opportunities for high-throughput materials discovery, particularly for high-entropy and compositionally complex ceramics with vast compositional spaces to explore.

**Data Availability** All data necessary to support the conclusions are provided within the paper and the cited literature. No original experimental data are presented in this perspective article.

**Acknowledgements** This work is supported by the Synthesis and Processing Science Program of the U.S. Department of Energy (DOE), Office of Science, Basic Energy Sciences (BES), Division of Materials Science and Engineering, under grant no. DE-SC0025255.

**Competing interests** The author declares that he has no competing interests.